# Optical cages made of graphitic frameworks

J. P. Walker and H. Grebel

The Electronic Imaging Center and the ECE department at NJIT, Newark, NJ 07102.
grebel@njit.edu

**Abstract:** In pursuit of infrared (IR) radiation absorbers, we examine quasi-crystal structures made of graphite wires. An array of graphitic cages and cage-within-cage, and whose overall dimensions is smaller than the radiation wavelength exhibit a flat absorption spectrum, A~0.83 between 10-30 microns and a quality loss factor of L~0.83 (L=A/Q, with Q – the quality factor). Simulations at microwave frequencies show multiple absorption lines. In the case of a cage within cage, energy is funneled towards the inner cage which result in a rather hot structure. Applications are envisioned as anti-fogging surfaces, EM shields and energy harvesting.

**Introduction:** There is a growing interest in perfect absorbers [1], which do not transmit, nor reflect electromagnetic (EM) radiation. In the past [2], we analyzed metallo-dielectric screens as highly absorbing structures in the near infrared (near IR). Here we extend this analysis to mid-IR wavelengths with graphitic wires.

Conductive-dielectric absorbers on a sub-wavelength scale may broadly fall into several categories: resonating structures such as meta-materials [3-5], interferometric devices [6-8] and lossy thin film or surface guides [9-10], most of which are quasi-two dimensional. Our structures are based on hollow frameworks, which are made of graphitic wire cages. Their overall thickness is less than a wavelength. Our simulations suggest that at the minimum, an array of quasi-crystal frameworks exhibits a large absorption coefficient (A=0.83) with a very large bandwidth of 20 microns, centered at $\lambda_0$=20 microns and are capable of trapping electromagnetic energy within them. A cage-within-cage framework exhibits a similar absorption coefficient (A~0.83), similar spectral width and funnels the radiation further towards its inner cage.

A metric, by which one compares all absorbers, regardless of their frequency of operation or their dielectric loss factor is defined as the absorption-bandwidth product, normalized by the center frequency of operation. The bandwidth-to-frequency ratio is just the inverse of the quality factor Q, thus, $L=A\Delta\lambda/\lambda=A/Q$. For example, an ultimate absorber, such as a frequency independent black body with an absorption coefficient of $A^{max}$=1, has very large bandwidth response and its center frequency is at the bandwidth center. In this case, the bandwidth is twice its center frequency: $Q^{max}$=0.5 and $L^{max}$=2. In the Yablonovitch limit [11], a frequency independent, mirror-clad, weakly absorbing film ($\alpha d$=0.02, n=3.5) in which the mirror suppresses the transmission, has it: $A=4n^2\alpha d/(1+4n^2\alpha d)$=0.5; Q~0.5; and L~1. Resonators allow both transmission and reflection modes and are made of metal films on a slightly lossy dielectric. They are typically narrow band. For example [3,4], an excellent absorption coefficient of A~0.99 with Q~25 at the microwave frequencies exhibits L~0.04. We emphasize though, that the effect in [3,4] is due to the lossy nature of the dielectric filler itself and less to do with the structured surface metal electrodes. We have shown that a bilayer structure can achieve 97% of absorption (A=0.97) in the near infrared without the necessity for a lossy dielectric material [12]. Here too, we do not include lossy dielectric material, though the conductor exhibits complex permittivity. Our graphitic icosahedral cages exhibit a Quality Loss Factor, L, where $L=A/Q\sim 0.83$, and $Q=1$ (since the bandwidth equals the center frequency).

A Faraday Cage [13], a hollow structure made of knitted conductive wires, shields its inner domain from external electromagnetic radiation through current loops at its surface. The openings in the wire mesh are typically very small compared to the effective radiation wavelength; thus, the radiation energy is dissipated at the cage's surfaces. Here, we concentrate on cages whose dimensions and openings are in the order of the radiation wavelength. The excited dipoles in the wire mesh generate an internal field, which is not fully frustrated as for its homogeneous surface counterpart [2]. Our quasi crystals structures take advantage of the interplay between the negative dielectric constant of graphite and its moderate conductive losses in the IR wavelength region. At longer wavelengths, the conductive wires behave as perfect electric conductors and if the icosahedral array is filled with a lossless dielectric, then the array becomes reflective.

Periodic metallo-dielectric structures (also known as screens, or metal meshes) have been studied in the past, and in particular in the long wavelength region – the wavelength region where the array pitch is of the order of, or smaller than the radiation wavelength [12, 14-16] – and thus, is above the diffraction region. Stacked periodic metal screens resemble photonic crystals with a large index of refraction ratio [17]. Conductive screens may be divided into two categories: inductive screens (conductive films with a periodic array of holes which portray a transmission band) and capacitive screens (the complementary structure where conductive structures are embedded in a dielectric and portray a reflection band). The screens exhibit negative index of refraction, or NIR. For inductive screens the NIR is exhibited throughout the wavelength band pass. For capacitive screens, the NIR region lies in the longer wavelength region beyond the reflection resonance. Our array of graphitic mesh structures may be viewed as capacitive screens.

**Simulations:** As indicated earlier, the structure takes advantage of the negative dielectric constant of graphite at the mid IR – the same effect that leads to surface plasmons polaritons (SPP) in that spectral range and the relatively low conduction losses [18]. The icosahedral edge and wire thickness varied. A CAD tool (Comsol) was employed in the analysis of the array. Periodic boundary conditions between the icosahedrons and perfect matching layers (PML) on top and bottom of the computation cell were used. Optimizations with edge-length to wire-thickness ratio may be made. The cell size was 10 x 10 x 100 micron$^3$. The incident intensity of 1 W is linearly polarized plane wave. The fluence of radiation is, therefore, 0.01 W/micron$^2$. Periodic boundary conditions are used with a perfect matched layer on the top and bottom of the cell. As for the thermal simulations, we used periodic boundary conditions around the edges of the unit cell. For example, we set the temperature along the x-direction, temperature(x,-g/2,z)=temperature(x,g/2,z), and along the y-direction, temperature(−g/2,y,z)= temperature(g/2,y,z)). The simulation works in one direction: the electromagnetic simulator affects the currents on the conductive layers, which result in heat. One may go one step further to assess how the resultant heat affects the electromagnetic absorption in a positive feedback manner; while it is worth exploring in future works, we found earlier that such approach only marginally affect the outcome [2]. The full-wave finite element solver with appropriate boundary conditions provides a complete solution to electromagnetic problems and was found in line with previous experimental results [19]

Plots of the intensity coefficients for the transmission, T, reflection, R, and total absorption A (which is defined here as, A=1-T-R) are provided in Fig. 1. The icosahedral array was illuminated by a plane wave, propagating along the negative z-direction with y-polarization. As shown in the figure, the absorption coefficient is substantial and flat across a wide range of wavelengths.

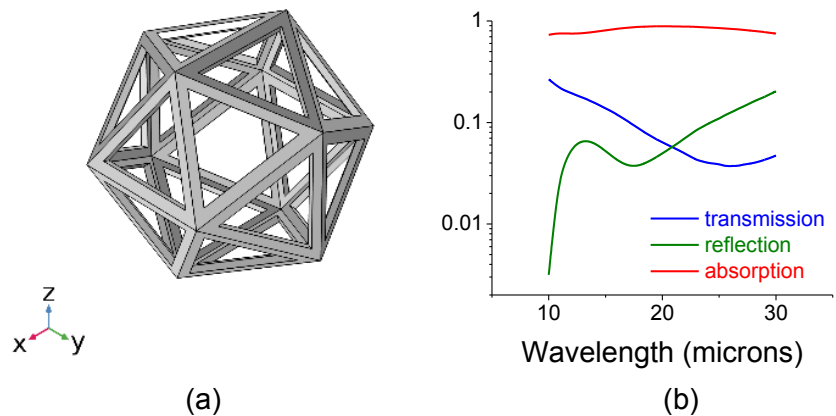

(a)             (b)

Fig. 1. (a) Icosahedral array made of a graphitis framework: the pitch is 10 microns and the edge is 6 microns. The wire thickness is 0.3 microns. (b) Coefficients for the transmission intensity, T, reflection intensity, R, and absorption A (defined as, A=1-T-R) as a function of wavelength in microns. The loss factor is L~0.83.

An aligned, cage-within-cage framework is shown in Fig. 2. Here, the enclosing (larger) icosahedron edge measures 6 microns and the enclosed (smaller) icosahedron edge measures 3 microns while the array pitch remains at 10 micron. The wire thickness are 0.3 and 0.15 microns for the outer and inner frameworks, respectively. Here, too, the absorption is flat across a large band of wavelengths.

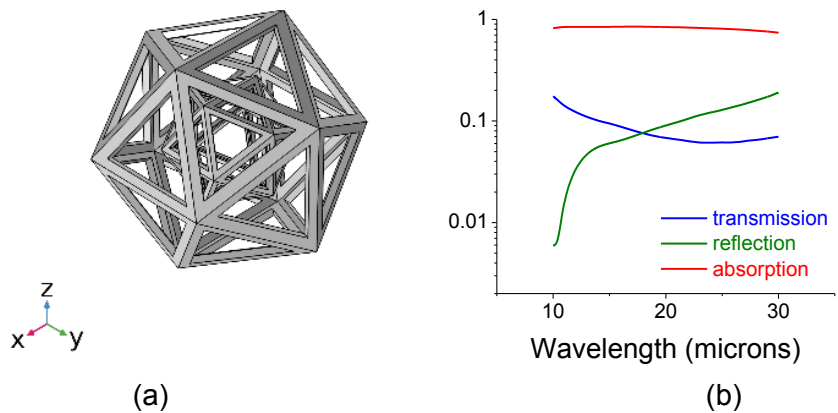

(a)             (b)

Fig. 2. (a) Aligned graphite-made cage-within-cage. (b) Coefficients for the transmission intensity, T, reflection intensity, R, and absorption A (defined as, A=1-T-R) as a function of wavelength in microns. The loss factor is L~0.83.

A simple graphitic cube is easier to fabricate than icosahedrons. In Fig. 3a,b we show the absorption of cubic wire structure made of graphite. As the cube opening becomes larger, the absorption peak is shifted towards the longer wavelengths (Fig. 3b). The peak absorptions are 0.85 and 0.89 for Fig. 3a and Fig. 3b, respectively. The FWHM at center wavelength equals to 15 microns is ~15 microns, namely, Q=1 and the quality loss factor is L~0.8.

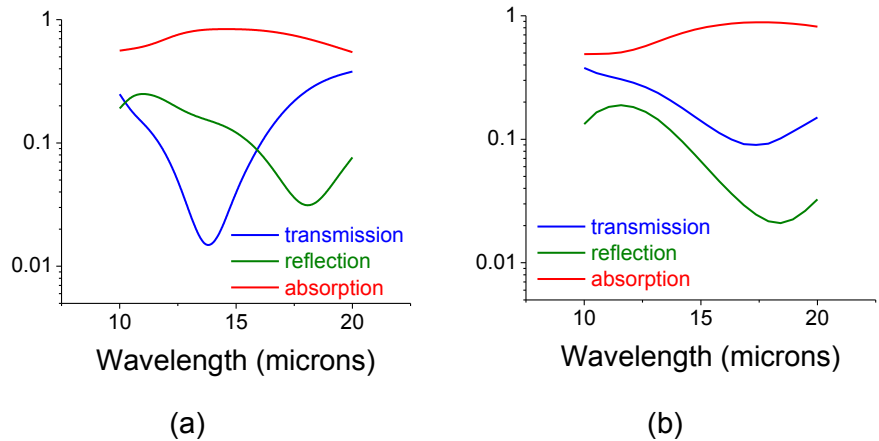

(a)  (b)

Fig. 3. (a) A simple graphitic cube whose dimensions are, edge length=6 microns and opening 4 microns exhibits a loss factor of L~0.8. (b) A graphitic cube whose dimensions are, length=6 microns and opening 4.8 microns also exhibits L~0.8.

**Microwave/RF:** the array is depicted in Figs. 4. Shown is a nested cubic array. The array of cubes exhibits large absorption in the cm wavelength range (GHz frequency range). At microwave frequencies, the graphite behaves essentially as a perfect conductor and the simulation was conducted for a perfect electric conductor (PEC) mesh of wires. The array was filled with an epoxy type filler having the following properties: relative permittivity=1; relative dielectric=4.5; conductivity 0.004 S/m. The filler radiation losses at microwave frequencies are, therefore, very small.

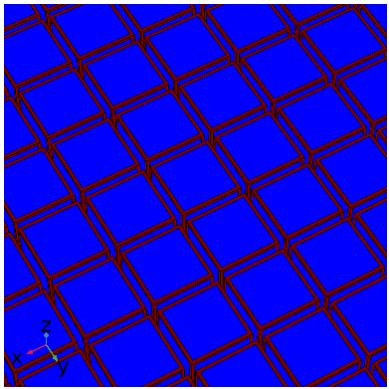 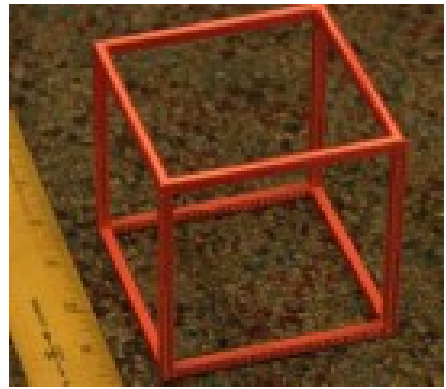

(a)  (b)

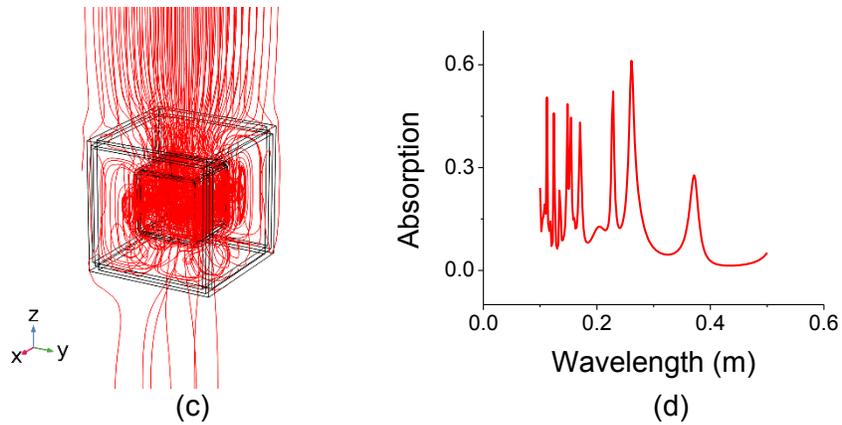

(c)                                      (d)

Fig. 4. Simple cubic array at microwave frequencies. (a) The array of cube wires. (b) A single cube made by additive manufacturing (namely, using a 3-D printer) and to be coated with a conductive film. The cube edge length is 9 cm. (c) Field lines for a box-within-box quasi-crystal array (one element is shown). The edge length of the inner box is 4.5 cm. Note the lensing effect to the radiation. (d) The absorption pattern for (c) with a low loss filler.

Capturing electromagnetic energy within the structures leads to heat and a quick rise in temperature. The temperature rise exhibits a linear trend over the time-scale studied; from 0 to 1 ns. Fig. 5 also indicates the funneling of thermal energy from the outer towards the inner cage. The electromagnetic intensity distribution is shown for the middle cross-section. Since the dielectric is made of air, the rise in temperature is seen on the conductive wires. Specifically, the inner cage becomes hotter. Both distributions were evaluated at a wavelength of 14.9 microns. The value of the trapped energy varies is a function of wavelength, though. For the cage-within-cage in the near infrared we found [2] that most of the trapped radiation energy at the peak absorption wavelength resides in between the cages as seems to be the case here.

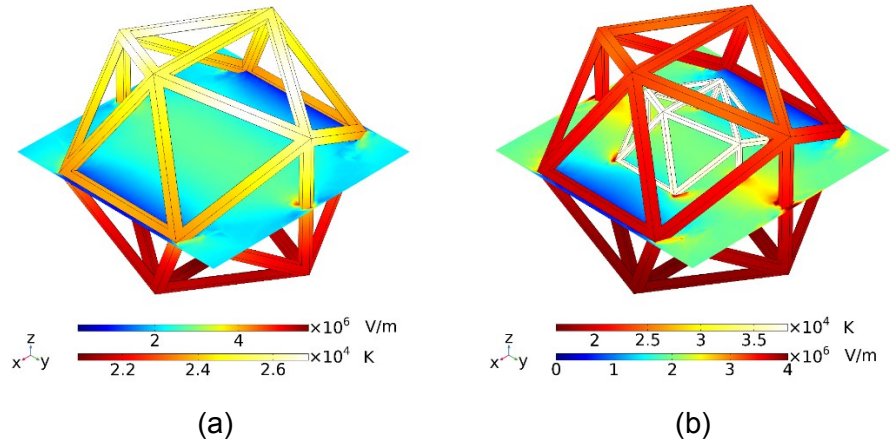

Fig. 5. Temperature and electromagnetic intensity distributions at λ=14.9 microns. (a) A single cage array and (b) a cage-within-cage array. The electromagnetic fluence is 0.01 W/micron$^2$. The simulation was captured after 1 ns. The electromagnetic intensity distribution is shown only for the middle cross-section whereas the temperature is developed on the conductive wires.

In summary, arrays of graphitic quasi-crystal wires were found to be a very effective radiation absorber in the IR and to lesser extent in the microwave wavelength range.